\documentstyle[amsfonts,aps,twocolumn]{revtex}
\input epsf

\begin{document}
\title{Resonant Control of Elastic Collisions in an Optically Trapped Fermi Gas of Atoms}
\author{T. Loftus \cite{adr1}, C. A. Regal, C. Ticknor, J. L. Bohn, and D. S. Jin \cite{adr1}}

\address{JILA, National Institute of Standards and Technology
and University of Colorado, Boulder, Colorado 80309}
\date{29 November, 2001}
\maketitle

\begin{abstract}
We have loaded an ultracold gas of fermionic atoms into a far off
resonance optical dipole trap and precisely controlled the spin
composition of the trapped gas.  We have measured a magnetic-field
Feshbach resonance between atoms in the two lowest energy
spin-states, $|9/2, -9/2\rangle$ and $|9/2, -7/2\rangle$. The
resonance peaks at a magnetic field of $201.5\pm1.4$ G and has a
width of $8.0\pm1.1$ G. Using this resonance we have changed the
elastic collision cross section in the gas by nearly 3 orders of
magnitude.
\\~\\PACS number(s): 34.50.-s, 32.80.Pj, 05.30.Fk\\~\\
\end{abstract}

Magnetic-field Feshbach resonances have recently emerged as a
powerful tool for studies of ultracold and quantum degenerate
gases\cite{Inou98,Cour98,Robe98,Vule99}. These resonances arise
when the collision energy of two free atoms coincides with that of
a quasi-bound molecular state\cite{Stwalley76,Ties92}. To access
the resonance a magnetic field is applied to tune relative
energies through the Zeeman effect. Measurements of Feshbach
resonances enable highly accurate determinations of molecular
interaction parameters\cite{Robe98,Robe01a,Chin00}. More
importantly, a Feshbach resonance gives the experimenter the
unique ability to control the interaction strength in a system.
The interactions between atoms in a gas can be adjusted from very
weak to very strong simply by controlling the magnitude of an
applied magnetic field.  In addition, a Feshbach resonance can be
used to control the sign of the s-wave scattering length, which
determines whether interactions between atoms are effectively
repulsive or attractive.  These abilities were recently exploited
in $^{85}$Rb experiments where a Feshbach resonance was used to
facilitate evaporative cooling to Bose-Einstein
condensation\cite{Corn00}. Furthermore, the resonance enabled a
detailed study of the collapse dynamics of Bose-Einstein
condensates with attractive interactions\cite{Robe01}. In a
quantum degenerate Fermi gas of atoms\cite{Dema99a,Trus01,Schr01}
a magnetic-field Feshbach resonance has the potential to play a
similar and equally valuable role in exploring radically different
interaction regimes. Additionally, a Feshbach resonance in a Fermi
gas offers the opportunity to access a predicted phase transition
to a novel superfluid state\cite{Stoo96,Holl01,Timm01}.

In this Letter we report the observation of a Feshbach resonance
between optically trapped fermionic atoms in two different
internal states. An optical dipole trap was used to confine atoms
in the two lowest energy Zeeman states of $^{40}$K,
$|F=9/2,m_F=-9/2\rangle$ and $|9/2,-7/2\rangle$, where $F$ is the
total atomic spin and $m_F$ is the magnetic quantum number. The
Feshbach resonance between atoms in these states drives a dramatic
change in the measured cross-dimensional rethermalization rate,
with a peak at an applied magnetic field of $201.5\pm1.4$ G, in
agreement with the theoretical prediction in Ref. \cite{Bohn00}.

The fermionic isotope $^{40}$K, with a nuclear spin of 4, has a
wealth of spin-states available for study (see Fig. 1). The
procedure for creating an ultracold gas of $^{40}$K atoms builds
upon our previous cooling and trapping
techniques\cite{Dema99a,Dema01a}. Atoms in the $|9/2,9/2\rangle$
and $|9/2,7/2\rangle$ states are first held in a magnetic trap and
cooled by forced evaporation. (Two spin components are used for
the evaporative cooling because fermions in the same internal
state stop colliding at ultralow temperatures\cite{Dema99b}.) The
ultracold gas is then loaded into a far-off resonance optical
dipole trap (FORT).  The FORT provides additional experimental
capabilities, such as the ability to confine atoms in any
combination of hyperfine spin-states with a trapping potential
that is independent of an atom's magnetic moment\cite{Sten99}.
This is particularly useful when working with fermionic atoms
since the realization of an interacting quantum Fermi gas requires
two (or more) components\cite{Dema01a,Gens01,Dema01b}. In
addition, optical trapping allows for the application of arbitrary
magnetic fields. Both capabilities are essential for accessing the
predicted Feshbach resonance in $^{40}$K since it occurs for
collisions between atoms in the ``strong-field-seeking"
$|9/2,-9/2\rangle$ and $|9/2,-7/2\rangle$ states, which cannot be
confined in a static magnetic trap.

The FORT consists of a single focused beam obtained from a
$\lambda$ = 1064 nm Nd:YAG laser. The trapping beam passes through
an acousto-optic modulator, which is used for intensity
stabilization and switching.  The beam is spatially filtered with
a single-mode optical fiber before being focused to a waist of
approximately 20 $\mu$m at the center of the magnetic trap.  To
load the atoms into the FORT we simply overlap the FORT beam with
the magnetic trap.  The FORT is switched on to 0.18 W in 40 ms
with a linear ramp.  The magnetic trap is then turned off quickly
($<$ 1 ms switching time) 10 ms later\cite{note1}.

Before imaging the atom cloud, the FORT light is suddenly switched
off and the gas is allowed to expand freely.  Following an
expansion time of typically 6 ms we take a picture of the gas
using resonant absorption imaging.  Number, temperature, and
center-of-mass position of the gas are extracted from the images.
As this imaging technique is destructive, all the measurements
described here involved repeated cycles of the experiment. To
observe the spin-composition of the gas, a magnetic field with a
strong vertical gradient is applied during the expansion from the
FORT\cite{Kett98,Dema01a}. Atoms in different spin states then
separate spatially via the Stern-Gerlach effect.  Figure 1 shows a
resonant absorption image of a gas that contains a mixture of all
10 Zeeman states in the $F=9/2$ hyperfine ground state of
$^{40}$K.

\begin{figure}
\begin{center}
\epsfxsize=2.0 truein \epsfbox{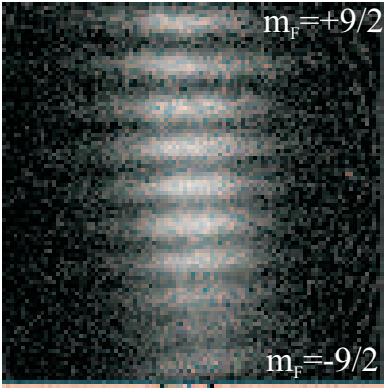}
\end{center}
\caption{A mixed spin-state gas released from the FORT.   Atoms in
the 10 Zeeman states of the $^{40}$K F=9/2 ground state are
simultaneously trapped in the FORT.  Before imaging, atoms in
different spin-states were separated with a Stern-Gerlach field
applied during 7.3 ms of expansion from the FORT. This picture
encompasses a 3.1 mm square area and is the sum of absorption
images taken with $\sigma^+$ and $\sigma^-$ light.  The mixture of
spin-states in the FORT was achieved by suddenly reversing the
direction of an applied magnetic field.
 } \label{spin state fig}
\end{figure}

To characterize the FORT strength, the radial frequency was
measured by modulating the optical power and monitoring the
resulting temperature of the gas. For a harmonic potential,
parametric heating occurs when the modulation frequency equals
twice the trap frequency.  In parametric heating measurements we
observed two transverse frequencies, vertical and horizontal,
which differed by a factor of 1.6.  We attribute this difference
to astigmatism or asymmetry in the FORT beam optics. For the 0.8 W
of optical power used for the measurements of the Feshbach
resonance, the lower, vertical frequency, $\nu_y$, was 1.2 kHz. In
the weaker axial direction the trap strength was measured by
exciting center-of-mass motion with a pulsed magnetic field
gradient. In the 0.8 W FORT the gas oscillated with a frequency of
15 Hz. However, strong damping of this axial motion after only a
couple cycles was indicative of strong anharmonicity in the axial
potential. As an operational measure of the trap depth we found
that the maximum temperature of a gas confined in the 0.8 W FORT
was 18 $\mu$K.

Exploitation of a Feshbach resonance between atoms in two internal
states requires precise control over the spin composition of the
gas.  Both elastic and inelastic collision rates can be strongly
affected by spin composition impurity. For example, if the elastic
collision cross-section between the two relevant states is tuned
to zero using the Feshbach resonance, even a small fraction of
atoms in a third spin state will dominate collisional interactions
in the gas and set a finite elastic collision rate. In addition,
atoms in other spin states can cause heating and number loss due
to inelastic processes such as spin-exchange collisions.

For measurements of the Feshbach resonance, the spin composition
of the gas was controlled using the following procedure.  Prior to
loading the FORT, the magnetically trapped $^{40}$K gas was
spin-polarized with all the atoms in the $|9/2,9/2\rangle$ state.
Atoms in the second spin-state used for the evaporative cooling,
$|9/2,7/2\rangle$, were selectively removed with a frequency-swept
microwave field, which drove transitions to an untrapped spin
state in the upper hyperfine ground state.  To ensure that the
remaining gas contained only atoms in the $|9/2,9/2\rangle$ state,
we applied a second frequency-swept microwave field that removed
any atoms in the $|9/2,5/2\rangle$ and $|9/2,3/2\rangle$ states.
The spin-polarized gas was then loaded into the FORT as described
above.  To preserve the spin composition we maintained an applied
magnetic field of at least 2 gauss while the atoms were in the
FORT.

After the gas is loaded into the FORT, we used adiabatic rapid
passage to create a mixture of atoms in the $|9/2,-9/2\rangle$ and
$|9/2,-7/2\rangle$ states.  The transfer used an applied
radio-frequency (rf) field and was performed in two steps. For
both steps a static, spatially uniform \cite{note2} magnetic field
of 28.5 G, generated by two of the magnetic trap coils, was
applied along the weaker axial direction of the FORT. In the first
step, the spin-polarized $|9/2,9/2\rangle$ gas was completely
transferred to the $|9/2,-9/2\rangle$ state with a 10 ms long
frequency sweep across all 9 Zeeman transitions. In the second
step, approximately 40$\%$ of the atoms were transferred to the
$|9/2,-7/2\rangle$ state by sweeping the rf frequency across the
$|9/2,-9/2\rangle$ to $|9/2,-7/2\rangle$ transition in 20 ms. With
the rf sweeps optimized for minimal transfer to other Zeeman
states, the fraction of atoms transferred to the
$|9/2,-7/2\rangle$ spin state was 36$\pm8\%$.

For our measurements of the Feshbach resonance the FORT power was
adiabatically increased from 0.18 W to 0.8 W and the magnetic
field was increased to between 160 G and 260 G.  The elastic
collision cross-section was then measured with a cross-dimensional
rethermalization technique\cite{Monr93}. The cloud was taken out
of thermal equilibrium by applying a parametric drive (modulating
the optical trap power) to preferentially heat the gas in the
vertical direction. The optical power in the FORT beam was
modulated for 3 to 5 ms with a 30$\%$ amplitude and a frequency of
2$\nu_y$. Following parametric heating, thermal relaxation of the
gas was observed in the time evolution of the rms cloud radii
$z_{rms}$, $y_{rms}$ in the axial and radial directions (see Fig.
2). A rethermalization time constant $\tau$ was extracted from an
exponential fit to the aspect ratio $z_{rms}$/$y_{rms}$ versus
time.

\begin{figure}
\begin{center}
\epsfxsize=4.0 truein \epsfbox{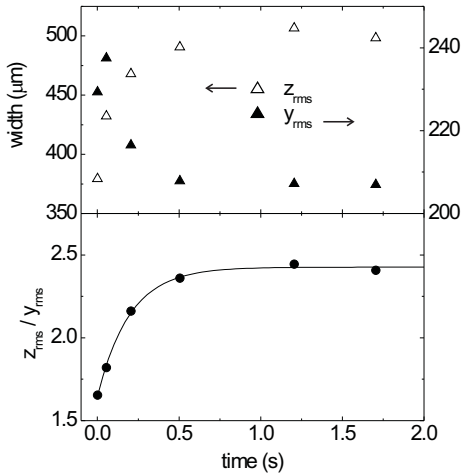}
\end{center}
\caption{Sample rethermalization data.  The rethermalization of
the optically trapped gas following modulation of the optical
power is shown for a 212.6 G magnetic field.  The cloud widths
$z_{rms}$ and $y_{rms}$ were obtained from gaussian fits to
absorption images taken 6 ms after the gas was released from the
FORT.  Following the parametric heating, energy in the gas is
redistributed by elastic collisions, and the ratio
$z_{rms}$/$y_{rms}$ approaches its equilibrium value.
 } \label{retherm fig}
\end{figure}

The elastic collision cross-section $\sigma$ was obtained from the
measured rethermalization time $\tau$ using $\frac{1}{\tau}$ =
$\frac{2}{\alpha} n \sigma v$. The constant $\alpha$ = 2.5 is the
calculated average number of binary \textit{s}-wave collisions
required for thermalization \cite{Dema99b}.  The mean relative
speed is given by $v=4\sqrt{\frac{k_BT}{\pi m}}$ and the density
overlap by $n=\frac{1}{N} \int\,n(\vec{r})^2 f (1-f)\,d^3\vec{r}$,
where $f$ is the fraction of atoms in the $|9/2,-7/2\rangle$
spin-state. The total number of atoms $N$, the temperature $T$,
and the density $n(\vec{r})$ were obtained from gaussian fits to
absorption images of equilibrated samples.

Figure 3 shows the elastic collision cross-section $\sigma$ as a
function of the applied magnetic field.  The Feshbach resonance is
clearly revealed as a change in $\sigma$ spanning nearly 3 orders
of magnitude.  At the peak of the resonance $\sigma$ is limited by
the finite temperature of the gas (T = 5 $\mu$K), while at the
zero of the resonance $\sigma$ is limited by a background
rethermalization rate.  For the data shown in Fig. 3 we have
subtracted this background rate, which was determined by measuring
the rethermalization rate in a spin-polarized $|9/2,-9/2\rangle$
gas and is consistent with spin-state contamination at the level
of 0.2$\%$. The error bars in Fig. 3 are dominated by
uncertainties in $\tau$ and the background rate. In addition,
$\sigma$ has an overall systematic uncertainty of approximately
$\pm50\%$ that comes from measurements of $N$. The magnetic field
values $B$ were calibrated using rf-driven spin-flip transitions
and have a systematic uncertainty of $\pm$ 0.5\%.

\begin{figure}
\begin{center}
\epsfxsize=3.5 truein \epsfbox{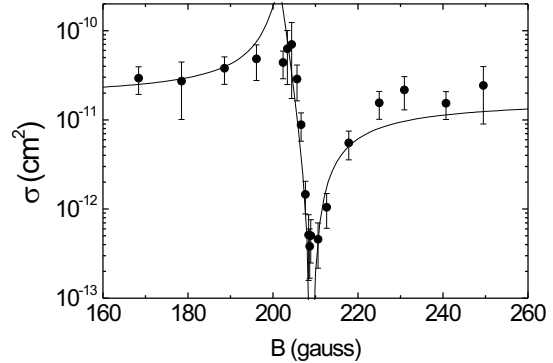}
\end{center}
\caption{Magnetic-field Feshbach resonance between $^{40}$K atoms
in the $|9/2,-9/2\rangle$ and $|9/2,-7/2\rangle$ states.  The
measured elastic collision cross-section $\sigma$, obtained from
measurements of the cross-dimensional rethermalization rate, is
plotted as a function of applied magnetic field $B$.  The data
were taken with $N$ = 8$\times$10$^6$, $T$ = 5 $\mu$K, and
\textit{f} = 0.36. The solid line is a fit to the data from which
we obtain cold collision parameters for potassium.
 } \label{resonance fig}
\end{figure}

The solid line in Fig. 3 is the best-fit theory curve, computed
from a multichannel scattering calculation whose details have been
discussed elsewhere \cite{Kmodel}.  In this calculation the
singlet and triplet scattering lengths $a_s$ and $a_t$ are
regarded as fully independent fitting parameters that uniquely
characterize the singlet and triplet potassium potentials.  The
model has been refined to incorporate both the accurate {\it ab
initio} van der Waals coefficient $C_6 = 3897 \pm 15$
\cite{Derevianko} and the correct number of bound states in the
$^{39}$K singlet and triplet potentials, $n_s = 86$ and $n_t =
27$, respectively \cite{Wang}. For each value of applied magnetic
field, thermally averaged cross sections are computed as described
in Ref. \cite{Dema99b}.  The reduced $\chi ^2$ is then minimized
with respect to $a_s$, $a_t$, and an overall scaling factor
$\epsilon$ that accounts for the systematic uncertainty in the
magnitude of the measured cross section.  The resulting best fit,
with $\chi ^2 = 1.06$, yields $a_s = 104.8 \pm 0.4$, $a_t = 174
\pm 7$, and $\epsilon = 1.30 $.  We find that $C_6$ does not play
a critical role in the shape or the position of the resonance, but
does affect the nominal value of $a_t$.  This uncertainty is
included in the quoted $a_t$ error bars. Performing the proper
mass scaling, the scattering lengths of $^{39}$K are found to be
$a_s = 139.4 \pm 0.7 a_o$ and $a_t = -37 \pm 6 a_0$, where $a_o$
is the Bohr radius. These results are in agreement with the values
of triplet scattering length reported in Ref. \cite{Wang} ($a_t =
-33 \pm 5 a_o$) and the singlet scattering length reported in Ref.
\cite{Kmodel} ($a_s =140^{+3}_{-6} a_o$).

In conclusion, we have demonstrated an optical trap for an
ultracold gas of fermionic atoms.  While the experiments described
here were carried out in the nondegenerate regime (at
approximately twice the Fermi temperature), future work will
explore an optically trapped quantum degenerate gas.  For example,
with an appropriate choice of FORT and magnetic trap parameters it
may be possible to achieve a high degree of degeneracy with an
adiabatic compression of the Fermi gas\cite{Vive01}. Further,
optical trapping has enabled us to observe a magnetic-field
Feshbach resonance in $^{40}K$.  This resonance is one of only a
handful of such resonances observed to date and furthermore is
unique in that it occurs for fermionic atoms in different internal
states. Future work will explore the utility of this tool for
controlling collisional interactions in a dilute Fermi gas.  We
have already seen density dependent heating and number loss around
the peak of the Feshbach resonance. Understanding these effects,
presumably due to inelastic collisions, will be important for
future applications. Of particular interest is the region where
the s-wave scattering length is negative, between the peak and the
minimum seen in our elastic collision cross-section measurements.
Here the interparticle interactions are effectively attractive and
could drive Cooper pairing of the fermionic
atoms\cite{Stoo96,Holl01,Timm01}.

This work is supported by the National Science Foundation, the
Office of Naval Research, and the National Institute of Standards
and Technology.  C. A. Regal acknowledges support from the Hertz
Foundation.


\begin{thebibliography}{10}
\bibitem{note1}
    We have observed additional heating of the gas for fast ($<$1 ms) switching of
    the FORT, but no dependence on the magnetic trap
    switching time or the overlap time.
\bibitem{note2}
    We estimate that the variation of the magnetic field across the axial extent of our largest
    cloud is less than 0.1\%.
\end{thebibliography}
\end{document}